\documentclass[aps,preprint,amsmath,amssymb,amsfonts,nofootinbib]{revtex4-1}
\usepackage{epsfig}
\usepackage{graphicx}
\usepackage{dcolumn}
\usepackage{bm}
\usepackage{amsthm}
\usepackage{amsmath}
\usepackage{color}
\usepackage{hyperref}

\newcommand{\beq}{\begin{equation}}
\newcommand{\eeq}{\end{equation}}
\newcommand{\bea}{\begin{eqnarray}}
\newcommand{\eea}{\end{eqnarray}}

\newcommand{\re}{\mathrm{Re}}

\begin{document}

\title{Liouville Conformal Field Theories in Higher Dimensions}
\author{Tom Levy and Yaron Oz}
\affiliation{Raymond and Beverly Sackler School of Physics and Astronomy, Tel Aviv University, Tel Aviv 69978, Israel}

\date{\today}
\begin{abstract}
We consider a generalization of the two-dimensional Liouville conformal field theory to any number of even dimensions. 
The theories consist of a log-correlated scalar field with a background ${\cal Q}$-curvature charge and an exponential  Liouville-type potential.
The theories are non-unitary and conformally invariant. They localize semiclassically on solutions that describe manifolds with a constant negative ${\cal Q}$-curvature. 
We show that $C_T$ is independent of the ${\cal Q}$-curvature charge and is the same as that of a higher derivative scalar theory.
We calculate the A-type Euler conformal anomaly of these theories. 
We study the correlation functions, derive an integral expression for them and calculate 
the three-point functions of light primary operators. The result is a higher-dimensional generalization 
of the two-dimensional DOZZ formula for the three-point function of such operators.

\end{abstract}


\maketitle

\section{Introduction}
Two-dimensional quantum Liouville theory has been a subject of much investigation since its 
first appearance in the study of non-critical string theory \cite{Polyakov:1981rd} 
(for reviews see e.g. \cite{Seiberg:1990eb,Teschner:2001rv,Nakayama:2004vk}). 
The theory provides a realization  of two-dimensional quantum gravity \cite{Knizhnik:1988ak,Distler:1988jt},
is an essential ingredient of many string theory backgrounds and has been related to certain $N=2$ SCFTs \cite{Alday:2009aq}.
As a conformal field theory (CFT) it is non-compact, thus 
the set of
Virasoro representations that make up its  space of states is continuous.

The aim of this work is to study  a generalization of the two-dimensional Liouville CFT to any number of even dimensions that
consists of a log-correlated scalar field with a background  ${\cal Q}$-curvature charge and an exponential  Liouville-type potential (for
an earlier work on the dynamics of the four-dimensional conformal factor see \cite{Antoniadis:1991fa}). 
Consider
an even-dimensional manifold $M$ of dimension $d$ without a boundary, equipped with a Euclidean signature metric $g_{ab}$.
The  action of the  higher-dimensional Liouville CFT  reads:
\begin{equation}
S_{L}(\phi,g) = \frac{d}{2\Omega_d(d-1)!}\int_{M}d^{d}x\sqrt{g}\left(\phi \mathcal{P}_g \phi+2Q \mathcal{Q}_g\phi+\frac{2}{d}\Omega_d(d-1)!\mu e^{db\phi}\right) \ .
\label{action}
\end{equation}
$\phi$ is a scalar field, $\mathcal{P}_g$   is the conformally covariant GJMS operator \cite{GJMS}:
\begin{equation}
\mathcal{P}_g =  (-\Box)^{\frac{d}{2}} + lower~ order \ ,
\end{equation} 
where $\Box = g^{ab}\nabla_a\nabla_b$ with $\nabla_a$ being the covariant derivative,
 and $\mathcal{Q}_g$ is the ${\cal Q}$-curvature  scalar \cite{Q}:
\begin{equation}
 {\cal Q} = \frac{1}{2(d-1)}(-\Box)^{\frac{d}{2}-1} R + ...  \ .
 \end{equation}
 
The dimensionless parameters in the action (\ref{action}) are the background charge  $Q$,  the cosmological constant $\mu$ and
$b$. $\Omega_d = \frac{2 \pi^{\frac{d+1}{2}}}{\Gamma[\frac{d+1}{2}]}$ is the surface volume of the $d$-dimensional sphere $S^d$.
When $d=2$ the action (\ref{action}) is that of the two-dimensional Liouville field theory.
When $\mu=0$, the action (\ref{action}) describes a  higher-dimensional Coulomb gas theory, and we will
denote this action by  $S_{C.G.}(\phi,g)$.
It appeared in \cite{Oz:2017ihc}
as part of a proposal for a field theory description of inertial range turbulence and the analysis of the A-type conformal anomaly.

The action (\ref{action}) defines non-unitary conformally invariant theories that localize semiclassically on solutions that describe manifolds with a constant negative ${\cal Q}$-curvature. 
We will show that $C_T$ is independent of the ${\cal Q}$-curvature charge and is the same as that of a higher derivative scalar theory
\cite{Osborn:2016bev}.
We will calculate the A-type Euler conformal anomaly of these theories. 
We will study the correlation functions, derive an integral expression for them and calculate 
the three-point functions of light primary operators. The result is a higher-dimensional generalization 
of the two-dimensional DOZZ formula for the three-point function of such operators  \cite{Dorn:1994xn,Zamolodchikov:1995aa}.

The paper is organized as follows.
In section 2 we will consider the classical higher-dimensional Liouville CFTs, verify their Weyl invariance, derive the field equations and define
the background  ${\cal Q}$-curvature charge.
In section 3 we will study the higher-dimensional Coulomb gas theory, the two-point function and the  quantum background  ${\cal Q}$-curvature charge,
 $C_T$ and the A-type conformal anomaly.
In section 4 we will analyze the Liouville correlation functions, 
derive an integral expression for them and calculate 
the three-point functions of light primary operators. Section 5 is devoted to a discussion and outlook.
In appendix A we briefly review the higher-dimensional  M\"{o}bius transformations that are used in section 4.

\section{Higher-Dimensional Liouville CFT}

\subsection{GJMS Operators and ${\cal Q}$-Curvature}

There are two objects in the action (\ref{action}) that play an important role in conformal geometry (for a review see e.g. \cite{conformal}). 
The first are  the conformally covariant GJMS operators  $\mathcal{P}_g$  \cite{GJMS}. For instance, in two and four dimensions they are the
Laplacian and the Paneitz operator \cite{Pan}, respectively:
\begin{equation}
\mathcal{P}_{d=2}   = -\Box,~~~~\mathcal{P}_{d=4} = \nabla_a\left(\nabla^a\nabla^b + 2 R^{ab} -\frac{2}{3}g^{ab}R\right)\nabla_b \ .
\end{equation}

The second object is the ${\cal Q}$-curvature  $\mathcal{Q}_g$ \cite{Q}, that takes in two and four dimensions the form:
\begin{equation}
{\cal Q}_{d=2} = \frac{1}{2} R,~~~~{\cal Q}_{d=4} = -\frac{1}{6}\left (\Box R + 3 R_{ab}R^{ab} - R^2 \right)  \ .
\end{equation} 
The integral of the ${\cal Q}$-curvature  on a Riemannian  manifold $M$ is an invariant of the conformal structure, but is not in general a topological invariant.
When $M$ is a conformally flat manifold, the  ${\cal Q}$-curvature is related to the Euler density $E_d$ and: 
\begin{equation}
\int_{M}d^{d}x\sqrt{g}\mathcal{Q}_{g} = \frac{1}{2}\Omega_{d}(d-1)!\chi(M) \ , \label{E}
\end{equation}
where $\chi(M)$ is the Euler characteristic of $M$.

\subsection{Classical Weyl Invariance}

Consider the Liouville CFT  defined by the action (\ref{action}).
Under a  Weyl transformation of the metric $g_{ab} \rightarrow e^{2 \sigma} g_{ab}$  the Liouville field $\phi$  transforms as:
\begin{equation}
\phi \to \phi-Q \sigma \ ,
\end{equation}
while the GJMS operator ${\cal P}_{g}$  transforms as:
\begin{equation}
{\cal P}_{e^{2 \sigma} g} = e^{- d \sigma} {\cal P}_{g} \ , \label{P}
\end{equation}
and the ${\cal Q}$-curvature  as:
\begin{equation}
{\cal{Q}}_{e^{2 \sigma} g} = e^{- d \sigma}\left({\cal{Q}}_{g} + {\cal P}_{g} \sigma \right) \ . \label{Qg}
\end{equation}
These transformations imply  that the action (\ref{action}) is classically Weyl invariant:
\begin{equation}
S_L(\phi-Q\sigma,e^{2\sigma}g)=S_L(\phi,g)-  S_{C.G.}(Q\sigma,g) \ ,
\end{equation}
for $Q = \frac{1}{b}$. This is the classical value of the background charge, and it will be modified by quantum corrections.
We denoted previously by $S_{C.G.}(\phi,g)$ the $\mu=0$ action (\ref{action}) that describes a  higher-dimensional Coulomb gas theory.

Equation (\ref{Qg}) can be written as:
\begin{equation}
{\cal P}_{g} \sigma + {\cal{Q}}_{g}  = {\cal{Q}}_{e^{2 \sigma} g} e^{d \sigma}  \ . \label{Qgg}
\end{equation}
If we take ${\cal{Q}}_{e^{2 \sigma} g} = {\cal Q}$ a real constant   then solutions $\sigma$ to equation (\ref{Qgg}) are answers to the question:
Given a manifold $M$ with a metric   $g_{ab}$, can we find  $\sigma$ such that 
under a Weyl transformation $g_{ab} \rightarrow e^{2 \sigma} g_{ab}$ we get a manifold that is conformally equivalent to $M$ and has
a constant ${\cal Q}$-curvature.
In two dimensions this means that we get a conformally equivalent surface with a constant scalar curvature. 
In higher dimensions the ${\cal Q}$-curvature does not determine the curvature tensor, however, the new metric with a constant
${\cal Q}$-curvature may have
special properties.

The field equations derived from the Liouville action (\ref{action}) for a rescaled Liouville field $\sigma =  b\phi$ take the form
(\ref{Qgg}) with a negative constant ${\cal Q}$-curvature:
\begin{equation}
{\cal P}_{g} \sigma + {\cal{Q}}_{g}  = -\Omega_d(d-1)!\mu b^2 e^{d\sigma} \ .
\end{equation}

\subsection{Background Charge}

For a conformally flat manifold with the topology of the sphere we get using (\ref{E}) 
for a constant shift of the field by $\phi_0$:
\begin{equation}
S_{C.G.}(\phi+\phi_0,g)=S_{C.G.}(\phi,g)+dQ\phi_0 \ .
\end{equation}
We will study these theories on the $d$-sphere $S^d$ and
the following discussion is a generalization of the two-dimensional analysis in \cite{Zamolodchikov:1995aa} to $d$-dimensions.
Since the sphere is conformally equivalent to flat space we can preform a (singular) Weyl transformation and work with a flat metric. 
The results boundary conditions for $\phi$ read:
\begin{equation}
 \phi(x)= -2Q\log\left(|x|\right)+O(1), \quad  |x|\to\infty \ , \label{BC}
 \end{equation} 
and this is called a background charge $-Q$ at infinity.

 When using a flat reference metric, one must regulate the region of integration and introduce boundary terms. We can define the action to be the large $R$ limit of:
   \begin{equation}
 S_L(\phi) = \frac{d}{2\Omega_{d}(d-1)!}\int_{B^d} d^{d}x\left(\phi (-\Box)^{\frac{d}{2}}\phi+\frac{2}{d}\Omega_{d}(d-1)!\mu e^{db\phi}\right)+\frac{dQ}{\Omega_{d-1}}\int_{\partial B^d}d^{d-1}\Omega\phi \ , \label{B}
 \end{equation}
 where $B^d$ is the $d$-dimensional ball of radius $R$ and $d^{d-1}\Omega$ is the volume element on its boundary $\partial B^d = S^{d-1}$. In writing down this action we have neglected boundary terms that are necessary in order to analyse the conformal boundary conditions for this theory \cite{Tom}, but are irrelevant to our analysis. In addition, this action needs to be regularized  in order to ensure its finiteness, and this is done by adding a constant term of the form $\mathcal{N}Q^2\log R$ where $\mathcal{N}$ is a suitable number. 
 
\subsection{The Semiclassical limit} 

The semiclassical limit of the theory is $b\to 0$.  In this limit it is convenient to work with the rescaled field $\phi_c= b\phi$ and the action 
(\ref{B}) reads:
  \begin{align}
 b^2S_L(\phi_c) &= \frac{d}{2\Omega_{d}(d-1)!}\int_{B} d^{d}x\left(\phi_c (-\Box)^{\frac{d}{2}}\phi_c+\frac{2}{d}\Omega_{d}(d-1)!\mu b^2 e^{d\phi_c}\right)
 \nonumber\\
 &\quad+\frac{d}{\Omega_{d-1}}\int_{\partial B}d^{d-1}\Omega\phi_c +O(b) \ . \label{BB}
 \end{align}
 The boundary condition (\ref{BC}) becomes:
 \begin{equation}
 \phi_c(x)= -2\log\left(|x|\right)+O(1), \quad  |x|\to\infty \ .
 \end{equation} 
The field equations that follow from (\ref{BB})  read:
\begin{equation}
(-\Box)^{\frac{d}{2}}\phi_{c} = -\Omega_{d}(d-1)!\mu b^2 e^{d\phi_{c}} \ ,
\label{FE}
\end{equation}
and this is equivalent to the equation:
\begin{equation}
\mathcal{Q}_{e^{2\phi_c}\delta_{ab}}=-\Omega_{d}(d-1)!\mu b^2 \ . \label{QC}
\end{equation}
Equation (\ref{QC})  describes a manifold with constant negative ${\cal Q}$-curvature.
Indeed, as discussed in a previous section, the semiclassical field equations resulting from the variation of (\ref{action}) on a manifold $M$ with a metric  $g_{ab}$ 
describe metrics on $M$ that are conformally 
equivalent to $g_{ab}$  with a constant negative ${\cal Q}$-curvature. 

\section{Higher-Dimensional  Coulomb Gas Theory}

\subsection{Two-point Function}

In Coulomb gas theory, i.e. $\mu =0$, the two-point correlation function of $\phi$ reads:
\begin{equation}
\left<\phi(x)\phi(0)\right> = \frac{2}{d}\log\left(\frac{L}{|x-y|}\right)+\dots \ ,
\end{equation}
where the computation is done using an IR regulator $L$ in the limit $L\to \infty$ and the dots indicate finite regulator dependent terms.
As in the two-dimensional case, we define vertex operators as: 
\begin{equation}
V_{\alpha}(x) = e^{d\alpha\phi(x)} \ , \label{vertex}
\end{equation}
which in  the free Coulomb gas theory are primary conformal operators of dimension:
\begin{equation}
\Delta_{\alpha} = d\alpha(Q-\alpha) \ . \label{del} 
\end{equation}
In the Liouville theory the result is the same, since we can compute the dimension of vertex operators by considering correlations in a state of our choice. By choosing a state in which $\phi\ll 0$ we can turn off the Liouville interaction potential  and reduce the calculation to the free field case.

In the Liouville CFT  we require that the interaction term has dimension $\Delta_{b}=d$, and thus using (\ref{del}) we get the quantum-corrected value of the background charge:
\begin{equation}
Q = b+\frac{1}{b} \ .
\end{equation}

\subsection{The Coefficient $C_T$}
The flat space stress-energy tensor is defined as the variation of the action with respect to the metric:
\begin{equation}
T_{ab} = \left.-\frac{2}{\sqrt{g}}\frac{\delta S}{\delta g^{ab}}\right|_{g_{ab}=\delta_{ab}} \ .
\end{equation}
As a result of the Weyl transformation law of the Liouville field and the quantum corrections to the dimension of the interaction term, the cosmological constant term in the Liouville action is not only Weyl invariant, it is in fact invariant under all variations of the metric. Therefore, the stress-energy tensors
obtained from the Liouville action and from  the Coulomb gas action are identical.

We can write the dependence of the flat space stress-energy tensor on the background charge in the following way:
\begin{equation}
T_{ab} = \left.T_{ab}\right|_{Q=0} + (-1)^{\frac{d}{2}} \frac{d Q}{\Omega_{d} (d-1)! (d-1)}\left(\partial_a \partial_b -\frac{1}{d}\delta_{ab}\Box\right)\Box^{\frac{d}{2}-1}\phi \ ,
\end{equation}
where $\left.T_{ab}\right|_{Q=0}$ is the stress energy tensor in the absence of a background charge, i.e. the part originating from the variation of $\mathcal{P}_{g}$.

Looking at the two-point function of the stress-energy tensor for Coulomb gas theory, and using the fact that the theory is Gaussian and the three-point function vanishes:
\begin{equation}
\left<T_{ab}(x)T_{cd}(0)\right> - \left.\left<T_{ab}(x)T_{cd}(0)\right>\right|_{Q=0} \propto \left(\partial_a \partial_b -\frac{\delta_{ab}}{d}\Box\right)\left(\partial_c \partial_d -\frac{\delta_{cd}}{d}\Box\right)\Box^{d-2}\left<\phi(x)\phi(0)\right> \ . \label{TT}
\end{equation}
The RHS vanishes (up to contact terms) for $d>2$. 

The coefficient $C_T$  is defined by:
\begin{equation}
\langle T_{ab}(x)T_{cd}(0) \rangle = \frac{C_T}{(x^2)^d} {\cal I}_{ab,cd}(x) \ , 
\end{equation}
where ${\cal I}_{ab,cd}(x)$ is the inversion tensor for traceless symmetric tensors.
The calculation (\ref{TT}) implies that for $d\neq 2$  the coefficient $C_T$ is independent of the background charge $Q$ and is the same as calculated for the higher derivative scalar theory in \cite{Osborn:2016bev}.

\subsection{A-type Conformal Anomaly}

Consider the quantum action of Coulomb gas theory:
\begin{equation}
W = -\log(Z) = \frac{1}{2}\log\left(\det\left(\frac{d}{\Omega_{d}(d-1)!}\sqrt{g}\mathcal{P}_g\right)\right)-\frac{1}{2}\left(\frac{d}{\Omega_{d}(d-1)!}\right)^2Q^2\mathcal{Q}_gD\mathcal{Q}_g \ ,
\end{equation}
where $D$ is the propagator of the theory:
\begin{equation}
(\sqrt{g}\mathcal{P}_g)_x D(x,y) = \frac{\Omega_d (d-1)!}{d}\delta^d (x,y) \ ,
\end{equation}
and we used the notation:
\begin{equation}
\mathcal{Q}_g D \mathcal{Q}_g \equiv \int d^{d}x d^{d}y \left(\sqrt{g}\mathcal{Q}_g\right)_x D(x,y)\left(\sqrt{g}\mathcal{Q}_g\right)_y \ .
\end{equation}

The A-type conformal anomaly coefficient $a$ is defined by:
\begin{equation}
\left<T^{a}_{a}\right>=\frac{1}{\sqrt{g}}\frac{\delta W}{\delta\sigma} = (-1)^{\frac{d}{2}+1}a E_{d} \ , \label{T}
\end{equation}
where $E_d$ is the Euler density normalized as:
\begin{equation}
\int_{S^d} \sqrt{g} d^d x E_d = \Omega_d d! \ ,
\end{equation}
and we work on a conformally flat space where all the Weyl invariant terms on the RHS of (\ref{T})
vanish. 
We get:
\begin{equation}
a = \frac{2}{\Omega_{d}(d!)^2}\int_{0}^{\frac{d}{2}} dt \prod_{i=0}^{\frac{d}{2}-1}\left(i^2-t^2\right)  +\frac{(-1)^{\frac{d}{2}}}{\Omega_{d}(d-1)!}Q^2  \ .
\label{a}
\end{equation}
The first term in (\ref{a}) has been calculated in \cite{Dowker:2010bu}.

\section{Liouville Correlation Functions}

\subsection{Correlation Functions}

We are interested in calculating correlation functions of the vertex operators (\ref{vertex}):
\begin{equation} \label{CorrelationFunc}
\left<V_{\alpha_1}(x_1)\cdots V_{\alpha_n}(x_n)\right> \equiv \int D\phi e^{-S_L}\prod_{i=1}^{n} e^{d\alpha_{i}\phi(x_i)} \ .
\end{equation}
By shifting $\phi \to \phi-\frac{\log \mu}{db}$ we get  using (\ref{E}) the KPZ scaling relation:
\begin{equation} \label{KPZ}
\left<V_{\alpha_1}\cdots V_{\alpha_n}\right> \propto \mu^{s},~~~~s = \frac{Q-\sum_{i} \alpha_i}{b} \ .
\end{equation}
Consider correlation functions a with fixed area $A = \int d^{d}xe^{db\phi}$. 
This is done by inserting the identity $1 = \int_{0}^{\infty} dA \delta\left(A-\int d^dx e^{db\phi}\right)$ into the Liouville functional. We get:
\begin{equation} \label{ALaplaceTrans}
\left< V_{\alpha_1}\cdots V_{\alpha_n}\right> = \int_{0}^{\infty}\frac{dA}{A}e^{-\mu A}\left<V_{\alpha_1}\cdots V_{\alpha_n}\right>_{A} \ ,
\end{equation}
where the fixed area correlation function is defined by:
\begin{equation} \label{FixedAreaPI}
\left<V_{\alpha_1}(x_1)\cdots V_{\alpha_n}(x_n)\right>_A = \int D\phi e^{-S_{C.G.}}A\delta\left(A-\int d^dx e^{db\phi}\right)\prod_{i=1}^{n} e^{d\alpha_{i}\phi(x_i)} \ .
\end{equation}

By shifting $\phi\to \phi + \frac{\log A}{db}$ one sees that the fixed area correlation functions satisfy the scaling relation:
\begin{equation}
\left<V_{\alpha_1}(x_1)\cdots V_{\alpha_n}(x_n)\right>_A = A^{-s}\left<V_{\alpha_1}(x_1)\cdots V_{\alpha_n}(x_n)\right>_{A=1} \ .
\end{equation}
Using the relation (\ref{ALaplaceTrans}) we find that for $\re{(s)}<0$:
\begin{equation}
\left< V_{\alpha_1}\cdots V_{\alpha_n}\right> = (\mu A)^{s}\Gamma(-s)\left<V_{\alpha_1}\cdots V_{\alpha_n}\right>_{A} \ .
\end{equation}

For $\re(s)\geq0$ the integral has a UV divergence at $A\to 0$. This corresponds to the fact that when $\re(s)\geq0$, as we will see later, there are no solutions to the classical equation of motion, i.e., there are no real saddle points. In this case the correlation function will include a non-universal part which is polynomial in $\mu$ and depends on  
a UV cutoff, and a universal cutoff independent part which is proportional to $(\mu A)^{s}\left<V_{a_1}\!\left(x_{1}\right)\cdots V_{a_n}\!\left(x_{n}\right)\right>_{A}$.

\subsection{Relation to a Free Field}

The KPZ scaling relation (\ref{KPZ}) shows that the correlation functions in Liouville theory are not analytic in $\mu$ and therefore we expect the
naive perturbation theory in $\mu$ to fail. This follows from the fact that by shifting the Liouville field we can always change the value of $\mu$ and therefore there is no sense in which we can consider it to be small.

We can separate the zero mode of the path integral over $\phi(x)$ from the non-zero mode $\phi_0$ and write $\phi(x)=\phi_0+\bar{\phi}(x)$. 
 As in the two-dimensional case (see e.g. \cite{Teschner:2001rv}), the Liouville measure factorizes in the following way:
\begin{equation}
D\phi e^{-S_L(\phi)} = d\phi_0 e^{-dQ\phi_0}[D\bar{\phi}]_{\mu,\phi_0} \ .
\end{equation}
Under translations of the zero mode the measure $d\phi_0$ is invariant and the measure over the non-zero mode satisfies   $[D\bar{\phi}]_{e^{-db\delta}\mu,\phi_0+\delta}=[D\bar{\phi}]_{\mu,\phi_0}$. In the limit $\phi_0\to -\infty$ the interaction term vanishes and the measure $[D\bar{\phi}]_{\mu,\phi_0}$ is asymptotic to the free Coulomb gas measure with corrections given by the interaction term:
\begin{equation} \label{AsympMeasure}
[D\bar{\phi}]_{\mu,\phi_0}\underset{\phi_0\to -\infty}{\sim} [D\bar{\phi}]_{free}e^{-S_{C.G.}\left(\bar{\phi}\right)}\sum_{n=0}^{\infty}\frac{(-\mu)^n e^{dnb\phi_0}}{n!}\left(\int d^dx e^{db\bar{\phi}}\right)^n \ .
\end{equation}

This is a power series in the translationally invariant small variable $\mu e^{db\phi_0}$. While the limit $\phi_0\to \infty$ of the integral over the zero mode is well behaved due to the presence of the Liouville-type interaction term, this is generally not the case for $\phi_0\to-\infty$. For the correlation functions (\ref{CorrelationFunc}) the leading  dependence in this limit is $e^{-dbs\phi_0}$ and we see that the integral has to be regularized for $\re(s)>0$. We can regularize it by subtracting the leading divergences, as given by the asymptotic behaviour (\ref{AsympMeasure}). One gets:

\begin{align}
\left<V_{\alpha_1}(x_1)\cdots V_{\alpha_N}(x_N)\right> &=\lim_{q_0\to -\infty}\left(\int^{\infty}_{q_0} d\phi_0 e^{-dbs\phi_0} \int [D\bar{\phi}]_{\mu,\phi_0}\prod_{i=1}^{N} e^{d\alpha_{i}\bar{\phi}(x_i)}\right. \nonumber \\
&\quad\left.-\sum_{n=0}^{\infty}\frac{(-\mu)^n}{n!}\frac{ e^{-d(s-n)bq_0}}{d(s-n)b}\int [D\bar{\phi}]_{free}e^{-S_{C.G.}(\bar{\phi})}\prod_{i=1}^{N} e^{d\alpha_{i}\bar{\phi}(x_i)}\left(\int d^dx e^{db\bar{\phi}}\right)^n\right) \ . \label{copo}
\end{align}
The correlation functions (\ref{copo}) have poles in the variable $s$ at values $s=n$. Denoting the residues $\mathcal{G}^{(n)}_{\alpha_1,\dots,\alpha_N}(x_1,\dots,x_N) = \;\underset{s=n}{\mathrm{Res}}\; \left<V_{\alpha_1}(x_1)\cdots V_{\alpha_N}(x_N)\right>$ we get:
\begin{align}
\mathcal{G}^{(n)}_{\alpha_1,\dots,\alpha_N} &= \frac{(-\mu)^n}{db n!}\int [D\bar{\phi}]_{free}e^{-S_{C.G.}(\bar{\phi})}\prod_{i=1}^{N} e^{d\alpha_{i}\bar{\phi}(x_i)}\left(\int d^dx e^{db\bar{\phi}}\right)^n\nonumber\\
&= \frac{(-\mu)^n}{n!}\int d^du_1\cdots d^du_n \left<V_{\alpha_1}(x_1)\cdots V_{\alpha_N}(x_N)V_{b}(u_1)\cdots V_{b}(u_n)\right>_{C.G.} \ .
\label{G}
\end{align}

In this equation $\left<\cdots\right>_{C.G.}$ denotes correlation functions in Coulomb gas theory, in which:
\begin{equation}
\left<V_{\alpha_1}(x_1)\cdots V_{\alpha_k}(x_k)\right>_{C.G.}  = \prod_{i<j} |x_i-x_j|^{-2d\alpha_i\alpha_j}, \quad \sum_{i=1}^{k}\alpha_k = Q \ .
\label{V}
\end{equation}

The correlation functions vanish unless $\sum_{i=1}^{k}\alpha_k = Q$.

\subsection{The Semiclassical Limit}

In this section we consider correlation functions of vertex operators :
\begin{equation}
\left<V_{\alpha_1}(x_1)\cdots V_{\alpha_n}(x_n)\right> \equiv \int D\phi_c e^{-S_L}\prod_{i=1}^{n} \exp\left(d\frac{\alpha_i}{b}\phi_{c}(x_i)\right) \ .
\label{integ}
\end{equation}
The following analysis is a generalization of the the two-dimensional case and follows the discussion in \cite{Zamolodchikov:1995aa,Harlow:2011ny}.

We wish to evaluate the integral (\ref{integ}) in the semiclassical limit $b\to 0$ using the saddle point approximation. 
The action (\ref{BB}) scales as $b^{-2}$, thus in order for a vertex operator insertion $V_{\alpha}$ in (\ref{integ}) 
to affect on the saddle points we require the scaling  $\alpha \sim b^{-1}$. 
We define $\alpha = \eta/b$, where we keep $\eta$ fixed when $b\to 0$. 
Such vertex operators define "heavy" operators whose dimensions read (\ref{del}):
\begin{equation}
\Delta = d\eta(1-\eta)/b^2,~~~ b\to 0 \ .
\end{equation}
"Light" operators are defined by vertex operators with $\alpha = b\sigma$ where $\sigma$ is kept fixed when $b\to 0$
and their dimension is $\Delta = d\sigma$ in the semiclassical limit.

The insertion of light operators can be quantified to lowest order in $b$ by a $b$-independent factor of $e^{d\sigma_i\phi_c(x_i)}$,
where  $\phi_c$ is the saddle point and hence it does not affect it.
On the other hand, an insertion of a heavy operator modifies the field equation (\ref{FE}):
\begin{equation} \label{SaddlePointEOM}
(-\Box)^{\frac{d}{2}}\phi_c = -\Omega_{d}(d-1)!\mu b^2 e^{d\phi_c} + \Omega_{d}(d-1)!\sum_{i}\eta_i\delta^{d}(x-x_i) \ .
\end{equation}
Assuming that in the neighbourhood of an operator insertion we can ignore the exponential term, one gets near a heavy operator the boundary condition: 
\begin{equation}
\phi_c(x) = -2\eta_i\log|x-x_i|+O(1), \quad x\to x_i \ .
\end{equation}

The physical metric $e^{2\phi_c(x)} \delta_{ab}$ in this region reads:
\begin{equation}
ds^2 = \frac{1}{r^{4\eta_i}} \left(dr^2+r^2 d\Omega^2_{d-1}\right) \ ,
\end{equation}
where $d\Omega^2_{d-1}$ is the metric on $S^{d-1}$
and the effect of a heavy operator can be interpreted as creating a conical singularity in the physical metric. Thus solving equation (\ref{SaddlePointEOM}) corresponds to finding conformally flat metrics of constant negative $\mathcal{Q}$-curvature on the sphere $S^d$ with the correct conical singularities.

Inserting the solution back into the field equations and requiring that the exponent is subleading one gets
the condition:
\begin{equation}
\re(\eta_i)<\frac{1}{2} \ . \label{cond}
\end{equation}
Condition (\ref{cond})  is called the Seiberg bound in  the two-dimensional case \cite{Seiberg:1990eb}.
It was interpreted as the non-existence  of local operators with $\re(\eta)>\frac{1}{2}$.  
Stated differently, $\alpha$ and $Q-\alpha$ correspond to the same quantum operator:
\begin{equation}
V_{Q-\alpha}=R(\alpha)V_{\alpha} \ ,
\end{equation}
where the relative scaling $R(\alpha)$ is called the reflection coefficient. When considering the semiclassical limit
we use out of these two operators the one that satisfies the Seiberg bound.

An additional constraint for real saddle points follows  from the Gauss-Bonnet-Chern theorem, by integrating (\ref{SaddlePointEOM}):
\begin{equation}
\sum_{i}\eta_i > 1 \ ,
\end{equation}
which  implies that there is no real saddle point  for the Liouville path integral with light operator insertions. 
When $\sum_i \eta_i <1$, we can consider the fixed area path integral (\ref{FixedAreaPI}), which still has a real saddle point. In the limit $b\to 0$, we can can fix the area $A = \int d^{d}x e^{d\phi_A}$, where $\phi_A = b\phi$, by using a Lagrange multiplier. This results in the following semiclassical equation of motion:
\begin{equation} \label{FASaddlePointEOM}
(-\Box)^{\frac{d}{2}}\phi_A = \frac{\Omega_{d}(d-1)!}{A}\left(1-\sum_i\eta_i\right)e^{d\phi_A} + \Omega_{d}(d-1)!\sum_{i}\eta_i\delta^{d}(x-x_i) \ .
\end{equation}
We see that in the case $\sum_i\eta_i<1$, the classical solutions correspond
to manifolds with positive constant ${\cal Q}$-curvature (that include the correct singularities) and finite area.

The action evaluated on a classical solution obeying our boundary conditions is divergent. 
In order to regularize it, we preform the action integral only over the part of the ball $B^d$ that excludes a ball $b_i$ of radius $\epsilon$ around each heavy operator insertion:
\begin{align}
 b^2 S_L &= \frac{d}{2\Omega_{d}(d-1)!}\int_{B\smallsetminus \cup_i d_i} d^{d}x\left(\phi_c (-\Box)^{\frac{d}{2}}\phi_c+\frac{2}{d}\Omega_{d}(d-1)!\mu b^2 e^{d\phi_c}\right) \nonumber\\
 &\quad+\frac{d}{\Omega_{d-1}}\int_{\partial B}d^{d-1}\Omega\phi_c  - \sum_i \frac{d\eta_i}{\Omega_{d-1}}\int_{\partial b_i}d^{d-1}\Omega\phi_c +O(b) \ .
\end{align}
The action is regularized by adding the field independent terms $\log R,\, \eta_i^2\log\epsilon$ multiplied by suitable numbers. The equations of motion for this action include both the equation of motion (\ref{SaddlePointEOM}) and the boundary conditions.

The leading exponential asymptotic in the limit $b\to 0$ for the correlation function of heavy and light operators is given by the semiclasssical expression:
\begin{equation} \label{SemiclassCorr}
\left<V_{\frac{\eta_1}{b}}(y_1)\cdots V_{\frac{\eta_n}{b}}(y_n)V_{b\sigma_1}(x_1)\cdots V_{b\sigma_m}(x_m)\right> \sim e^{-S_L(\phi_\eta)}\prod_{i=1}^{m} e^{d\sigma_i\phi_{\eta}(x_i)} \ ,
\end{equation}
where $\phi_{\eta}$ is the solution of the equation of motion obeying the correct boundary conditions. This formula includes effects that are $O(b^{-2})$ in the exponent exactly, while $O(b^0)$ effects are included only if they depend on the positions or conformal dimensions of the light operators.   In general there will be more than one solution, and the right hand side will include a sum, or an integral, over the solutions.

\subsection{Three-Point Functions of Light Primary Operators}

In a conformal field theory, the three-point function of primary operators is determined up to a constant by conformal invariance. In particular we have:
\begin{equation}
\left<V_{\alpha_1}(x_1)V_{\alpha_2}(x_2)V_{\alpha_3}(x_3)\right> = \frac{C(\alpha_1,\alpha_2,\alpha_3)}{|x_{12}|^{\Delta_1+\Delta_2-\Delta_3}|x_{13}|^{\Delta_1+\Delta_3-\Delta_2}|x_{23}|^{\Delta_2+\Delta_3-\Delta_1}} \ ,
\end{equation}
where the function $C(\alpha_1,\alpha_2,\alpha_3)$ specifies the structure constants of Liouville field theory.

We now consider the case where all three operators are light and therefore we need to examine the fixed area correlation function. The relevant solution to the fixed area equation of motion is the sphere metric of area $A$ (\ref{FASaddlePointEOM}):
\begin{equation}
\phi_{A}(x) = -\log\left(\frac{|x|^2+1}{2}\right)+\frac{1}{d}\log\left(\frac{A}{\Omega_d}\right) \ .
\end{equation}

We have to integrate over all solutions related to this one by conformal mappings, as follows from the conformal invariance of the problem. According to Liouville's theorem, all conformal mappings on a domain of $\mathbb{R}^{d}$ for $d>2$ are a composition of translations, inversions, dilations and orthogonal transformation, i.e. are higher-dimensional M\"{o}bius transformations. We describe these transformations using $2\times 2$-matrices with entries in the Clifford algebra $C_{d-1} = \mathrm{C}\ell_{0,d-1}(\mathbf{R})$, as detailed in appendix \ref{HDMobius}. In this formalism, higher-dimensional M\"{o}bius transformations can be written as $x \to (\alpha x+\beta)(\gamma x +\delta)^{-1}$ where $\alpha,\beta,\gamma,\delta \in \mathit{\Gamma}_{d-1}\cup \{0\}$, $\alpha\beta^{*},\gamma\delta^{*},\gamma^*\alpha,\delta^*\beta\in\mathbb{R}^{d}$ and $\alpha\delta^*-\beta\gamma^*=1$. This conformal mapping introduces a Weyl transformation with $\sigma = 2\log\left(|\gamma x +\delta|\right)$.

We can now write the general  M\"{o}bius transformation of the saddle point:
\begin{equation} \label{phiMobius}
\phi_{A}(x) = -\log\left(\frac{|\alpha x+\beta|^2+|\gamma x+\delta|^2}{2}\right)+\frac{1}{d}\log\left(\frac{A}{\Omega_d}\right) \ ,
\end{equation}
We now see that the moduli space of saddle points is given by $SL(2,C_{d-1})/SU(2,C_{d-1})$, because elements of the subgroup $SU(2,C_{d-1})$ of $SL(2,C_{d-1})$ leave the solution (\ref{phiMobius}) fixed.

Note, that in equation (\ref{SemiclassCorr}) when there are no heavy operators included all effects of operator insertions are $O(b^0)$ in the exponent. If we wanted to include all effects of this order, we would need to renormalize the functional determinant $\det\left(\frac{\delta^2 S_{C.G.}(\phi_A)}{\delta\phi^2}\right)$ and would also need the Jacobian for changing the integral over $\phi_A$ to an integral over $\alpha,\beta,\gamma,\delta$. We explicitly include $O(b^0)$ terms in the action, but represent the functional determinant and Jacobian as a $b$-dependent factor $\widehat{\mathcal{A}}(b)$ whose logarithm is at most $O(\log b)$ \cite{Harlow:2011ny}. Note, that it is independent of $\sigma_i$ since neither effect is affected by light operator insertions. 

We can now write:
\begin{equation} \label{Light3Point}
\left<V_{b\sigma_1}(x_1)V_{b\sigma_2}(x_2)V_{b\sigma_3}(x_3)\right>_{A}\approx \widehat{\mathcal{A}}(b)e^{-S_{C.G.}(\phi_A)}\int d\mu(\alpha,\beta,\gamma,\delta)\prod_{i=1}^{3}e^{d\sigma_{i}\phi_{A}(x_i)} \ ,
\end{equation}
where $d\mu(\alpha,\beta,\gamma,\delta)$ is the invariant measure on $SL(2,C_{d-1})$. The Coulomb gas action is given by :
\begin{align}
 S_{C.G.} &= \frac{1}{b^2}\frac{d}{2\Omega_{d}(d-1)!}\int_{B} d^{d}x\;\phi_A (-\Box)^{\frac{d}{2}}\phi_A+\left(\frac{1}{b^2}+1\right)\frac{d}{\Omega_{d-1}}\int_{\partial B}d^{d-1}\Omega\phi_A \ .
 \end{align}
 
 Evaluating this action for the solution (\ref{phiMobius}) we get:
 \begin{equation}
 S_{C.G.}(\phi_A) = \frac{1}{b^2}\left[S_{Bulk}+\log\left(2^d\frac{A}{\Omega^{d}}\right)\right]+\log\left(2^{d}\frac{A}{\Omega^{d}}\right)+O(b^2) \ .
 \end{equation}
 
 Here the constant $S_{Bulk}$ is given by regularizing (i.e. taking the finite part of) the large $R$ limit of the integral:
 \begin{equation}
 S_{Bulk} = \frac{d\Omega_{d-1}}{2\Omega_d(d-1)!}\int_{0}^{R}dr\,r^{d-1}\left(\Box_r^{\frac{d}{4}}\log\left(1+r^2\right)\right)^2 \ ,
 \end{equation}
 where $\Box_r = \frac{1}{r^{d-1}}\partial_r\left(r^{d-1}\partial_r\right)$ is the radial part of the Laplacian and 
 $\Box_r^{\frac{d}{4}}$ stands for  $\partial_r\left(\Box_r^{\frac{d-2}{4}}\right)$ in the case of odd $\frac{d}{2}$.
 
 Using the $SL(2,C_{d-1})$ transformation properties of the integral (\ref{Light3Point}) we can write \cite{Harlow:2011ny}:
 \begin{equation}
 \int d\mu \prod_{i=1}^{3}e^{d\sigma_{i}\phi_{A}(x_i)} = \left(2^d\frac{A}{\Omega^{d}}\right)^{\sum_i \sigma_i} |x_{12}|^{d(\sigma_3-\sigma_1-\sigma_2)}|x_{23}|^{d(\sigma_1-\sigma_2-\sigma_3)}|x_{13}|^{d(\sigma_2-\sigma_1-\sigma_3)}\hat{I}(\sigma_1,\sigma_2,\sigma_3) \ ,
 \end{equation}
 where:
  \begin{equation}
 \hat{I}(\sigma_1,\sigma_2,\sigma_3) = \int \frac{d\mu(\alpha,\beta,\gamma,\delta)}{\left(|\beta|^2+|\delta|^2\right)^{d\sigma_1}\left(|\alpha+\beta|^2+|\gamma+\delta|^2\right)^{d\sigma_2}\left(|\alpha|^2+|\gamma|^2\right)^{d\sigma_3}} \ .
 \end{equation}
 
 This integral is invariant under the $SU(2,C_{d-1})$ subgroup of $SL(2,C_{d-1})$. We can now parametrize $SL(2,C_{d-1})$ elements by a unitary matrix times an upper triangular matrix (i.e. a composition of dilatation and translation). The relevant integral is therefore:
  \begin{equation}
 I(\sigma_1,\sigma_2,\sigma_3) =\int_{0}^{\infty}\frac{d\lambda}{\lambda}\lambda^{d(\sigma_1+\sigma_2-\sigma_3)}\int d^{d}w \frac{1}{\left(|w|^2+1\right)^{d\sigma_1}\left(\left|w+\lambda\right|^2+1\right)^{d\sigma_2}} \ .
 \end{equation} 
  In changing coordinates from $\alpha,\beta,\gamma,\delta$ to $\lambda,w$ we get a $b$-independent Jacobian which we can ignore by replacing $\widehat{\mathcal{A}}(b)$ by a new factor $\mathcal{A}(b)$. This integral can be evaluated explicitly with the result \cite{Harlow:2011ny}:
  \begin{equation}
I(\sigma_1,\sigma_2,\sigma_3) = \frac{\pi^{\frac{d}{2}}}{2}\frac{\Gamma\left(\frac{d}{2}(\sigma_1+\sigma_2+\sigma_3-1)\right)\Gamma\left(\frac{d}{2}(\sigma_2+\sigma_3-\sigma_1)\right)\Gamma\left(\frac{d}{2}(\sigma_3+\sigma_1-\sigma_2)\right)\Gamma\left(\frac{d}{2}(\sigma_1+\sigma_2-\sigma_3)\right)}{\Gamma(d\sigma_1)\Gamma(d\sigma_2)\Gamma(d\sigma_3)} \ .
\end{equation}

 Finally, we find the semiclassical result for the structure constants of light operators:
\begin{align}
C(b\sigma_i)&\approx \frac{\pi^{\frac{d}{2}}}{2}\mathcal{A}(b)\left(2^d\frac{A}{\Omega^{d}}\right)^{\sum_i \sigma_i-1/b^2-1}e^{-S_{bulk}/b^2} \nonumber\\
&\quad\times \frac{\Gamma\left(\frac{d}{2}(\sigma_1+\sigma_2+\sigma_3-1)\right)\Gamma\left(\frac{d}{2}(\sigma_2+\sigma_3-\sigma_1)\right)\Gamma\left(\frac{d}{2}(\sigma_3+\sigma_1-\sigma_2)\right)\Gamma\left(\frac{d}{2}(\sigma_1+\sigma_2-\sigma_3)\right)}{\Gamma(d\sigma_1)\Gamma(d\sigma_2)\Gamma(d\sigma_3)}
\ .
\end{align}
This is the higher-dimensional generalization 
of the two-dimensional DOZZ formula for the three-point function of light primary operators \cite{Dorn:1994xn,Zamolodchikov:1995aa}.

\section{Discussion and Outlook}

In this work we initiated the study of a higher-dimensional generalization of the two-dimensional Liouville CFT
that consists of a log-correlated scalar field with a background ${\cal Q}$-curvature charge and an exponential  Liouville-type potential.
There are many interesting classical and quantum aspects of these theories that deserve further study.
Classically, the solutions to the field equations describe manifolds with a constant negative ${\cal Q}$-curvature. 
The space of solutions to this mathematical problem is not known in more than two dimensions and  it corresponds to a higher-dimensional
uniformization-like problem.

Quantum mechanically, it is quite possible that these theories can be solved once the three-point function
is calculated exactly. We calculated it for the special case of three light primary operators.
Performing the integral expression (\ref{G}), (\ref{V}) in general and deriving the exact formula for the three-point functions 
is an interesting and challenging problem. 

Another interesting direction to follow is to study these theories in a Lorentzian siganture.  Being non-unitary and higher derivative
theories it is not clear whether they can be defined and analyzed consistently in such a signature.

Much research on adding boundaries to CFTs  revealed a rather rich structure  in diverse dimensions.
It would be interesting to study the higher-dimensional Liouville CFTs in the presence of boundaries.
One needs to formulate the consistent boundary conditions and study issues like boundary operators,  correlation
functions and boundary anomalies \cite{Tom}. One can also consider the odd-dimensional bulk case with an even-dimensional
boundary, where the GJMS-type operators are pseudo-differential \cite{Oz:2017ihc}.

One can add fermionic degrees of freedom to the higher-dimensional Liouville CFTs and, as in the two-dimensional case,
construct and  study
supersymmetric versions of them.
Finally, it would be interesting to explore the possible role of the  higher-dimensional Liouville CFTs in the study of higher-dimensional random 
geometry, as well as a possible generalization of the AGT relation \cite{Alday:2009aq}.

\section*{Acknowledgements}
We would like to thank A. Schwimmer and S. Theisen for a discussion and I. Antoniadis for pointing out reference \cite{Antoniadis:1991fa} to us.
This work  is supported in part by the I-CORE program of Planning and Budgeting Committee (grant number 1937/12), the US-Israel Binational Science Foundation, GIF and the ISF Center of Excellence.

\newpage

\appendix

\section{Higher-Dimensional M\"{o}bius Transformation} \label{HDMobius}

This appendix is based on \cite{Waterman:1993}.

\subsection{Clifford Algebras}
The Clifford algebra $C_{n} = \mathrm{C}\ell_{0,n}(\mathbf{R})$ is the real associative algebra generated by $n$ elements $i_1,i_2,\dots,i_n$ subject to the relations $i_hi_k =  -i_ki_h, h\neq k, \; i_k^2=-1$ and no others. Every element of $C_n$ can be expressed uniquely in the form $a = \sum a_I I$ where the sum is preformed over all products $I =i_{v_1}i_{v_2}\cdots i_{v_k}$ with $1\leq v_1<v_2<\dots<v_k\leq n$ and $a_I\in\mathbb{R}$. The null product is permitted and identified with the real number $1$. $C_n$ is therefore a real vector space of dimension $2^n$ and one can identify $C_0$ with $\mathbb{R}$, $C_1$ with $\mathbb{C}$ and $C_2$ with the quaternions $\mathbb{H}$.
We give $C_n$ the Euclidean norm so that if $a = \sum a_II$ then $|a|^2 = \sum a_I^2$. There are three involutions of $C_n$:

\begin{enumerate}
\item $^*$: replace each $I = i_{v_1}i_{v_2}\cdots i_{v_k}$  with $i_{v_k}\cdots i_{v_2}i_{v_1}$. It determines an anti-automorphism of $C_n$: $(ab)^*=b^*a^*$.
\item $'$: replace each $i_k$  with $-i_k$. It determines an automorphism of $C_n$: $(ab)'=a'b'$.
\item $\bar{\,}$: $\bar{a}=(a')^*=(a^*)'$. It determines an anti-automorphism of $C_n$.
\end{enumerate}

Clifford numbers of the form $x = x_0 + x_1i_1+\dots+x_ni_n$ are called \textit{vectors}. They form an $(n+1)$-dimensional subspace which we identify with $\mathbb{R}^{n+1}$. For vectors $x^*=x$ and thus $x' =\bar{x}$. Further, $x\bar{x}=\bar{x}x = |x|^2$ so non-zero vectors are invertible  with $x^{-1} = \bar{x}/|x|^2$. Thus products of non-zero vectors are   invertible and form a multiplicative group, the \textit{Clifford group} $\mathit{\Gamma}_n$.

We have the following important property: If $a\in\mathit{\Gamma}_n$ and $x\in\mathbb{R}^{n+1}$ then $axa'^{-1}\in\mathbb{R}^{n+1}$ and the mapping $\rho_a(x)= axa'^{-1}$ is orthogonal. Furthermore, the map $\phi:\mathit{\Gamma}_n\to O(n+1)$ given by $\phi(a) = \rho_a$ is onto $SO(n+1)$ with kernel $\mathbb{R}\smallsetminus\{0\}$.

\subsection{Clifford Matrices}

We now consider Clifford matrices $T = \left(\begin{array}{cc} a & b \\
c & d \end{array}\right)$ with elements in $\mathit{\Gamma}_n\cup{0}$. Each such matrix is identified with the map $T:\hat{\mathbb{R}}^{n+1}\to\hat{\mathbb{R}}^{n+1}$ (where $\hat{\mathbb{R}}^{n+1}=\mathbb{R}^{n+1} \cup\{\infty\}$ is the one-point compactification of $\mathbb{R}^{n+1}$) given by:

\begin{equation}
Tx = (ax+b)(cx+d)^{-1}
\end{equation}

The group of invertible $2\times 2$ Clifford matrices is given by:

\begin{equation}
GL(2,C_n) = \left\lbrace  \left(\begin{array}{cc} a & b \\
c & d \end{array}\right): ab^{*},cd^{*},c^*a,d^*b\in\mathbb{R}^{d},\; ad^*-bc^*\in\mathbb{R}\smallsetminus\{0\} \right\rbrace
\end{equation}

The quantity $\Delta(T) = ad^*-bc^*$ is called the \textit{pseudo-determinant}. The inverse matrix is given by  $T^{-1} = \Delta(T)^{-1} \left(\begin{array}{cc} d^* & -b^* \\
-c^* & a^* \end{array}\right)$. A Clifford matrix induces a bijective mapping $\hat{\mathbb{R}}^{n+1}\to\hat{\mathbb{R}}^{n+1}$ if and only of it is in $GL(2,C_n)$. 

\bigskip
\bigskip

Each $T\in GL(2,C_n)$ is a composition of:

\begin{itemize}

\item Translation: $\quad\quad \left(\begin{array}{cc} 1 & \mu \\
0 & 1 \end{array}\right), \quad\quad x\to x+\mu,\quad \quad \mu\in \mathbb{R}^{n+1}$
\item Inversion: $\quad\quad \left(\begin{array}{cc} 0 & 1 \\
-1 & 0 \end{array}\right), \quad\quad x\to -x^{-1}=-\frac{\bar{x}}{|x|^2}$
\item Dilation: $\quad\quad \left(\begin{array}{cc} \sqrt{\lambda} & 0 \\
0 & 1/\sqrt{\lambda} \end{array}\right), \quad\quad x\to\lambda x,\quad \quad \lambda\in \mathbb{R}^{+}$
\item Trivial: $\quad\quad \left(\begin{array}{cc} \alpha & 0 \\
0 & \alpha \end{array}\right), \quad\quad x\to x,\quad \quad \alpha\in \mathbb{R}\smallsetminus\{0\}$
\item Special orthogonal: $\quad\quad \left(\begin{array}{cc} a & 0 \\
0 & a' \end{array}\right), \quad\quad x\to axa^*,\quad \quad a\in \mathit{\Gamma}_n,|a|=1$
\item Reflection: $\quad\quad \left(\begin{array}{cc} 1 & 0 \\
0 & -1 \end{array}\right), \quad\quad x\to -x$
\end{itemize}
Further, $T$ is orientation preserving if and only if $\Delta(T)>0$.

\bigskip\bigskip

We define $PGL(2,C_n) = GL(2,C_n)/(\mathbb{R}\smallsetminus\{0\})$, $SL(2,C_n) = \left\lbrace T\in GL(2,C_n):\Delta(T)=1\right\rbrace$, $PSL(2,C_n)=SL(2,C_n)/\mathbb{Z}_2$. Then $PGL(2,C_n)$ is isomorphic to $GM(n+1)$, the full group of M\"{o}bius transformations of $\hat{\mathbb{R}}^{n+1}$ and $PSL(2,C_n)$ is isomorphic to $M(n+1)$, the group of orientation preserving M\"{o}bius transformations of $\hat{\mathbb{R}}^{n+1}$.

The linear distortion of $T\in GL(2,C_n)$ is $|T'(x)| = \frac{|\Delta|}{|cx+d|^2}$.

Finally, we define the group:

\begin{equation}
SU(2,C_n) = \left\lbrace  \left(\begin{array}{cc} a & b \\
-b' & a' \end{array}\right)\in SL(2,C_n)\right\rbrace 
\end{equation}

For each Clifford matrix $T = \left(\begin{array}{cc} a & b \\
c & d \end{array}\right)$ we define $||T||^2 = |a|^2+|b|^2+|c|^2+|d|^2$. Then for all $U\in SU(2,C_n)$ we have $||UT||=||TU||=||T||$.

\end{document}